\documentclass[10pt]{article}
\usepackage[dvips]{epsfig}
\usepackage{epsfig,psfig,latexsym,citesort}
\usepackage{amssymb,amsmath}
\usepackage{natbib}

\input{amssym.def}
\input{amssym}

\newtheorem{lemma}{Lemma}
\newcommand{\qed}{~$\vrule width.15cm height.2cm depth0cm$ \medbreak}
\newenvironment{proof}{\noindent{\bf Proof: }}{\qed}

\newcommand {\mm}[1]{\ifmmode{#1}\else{\mbox{\(#1\)}}\fi}

\newcommand{\real}{\mm{{\Bbb R}}}

\newcommand{\utwi}[1]{\mbox{\boldmath $ #1$}}
\newcommand{\ba}{{\utwi{a}}}

\newcommand{\bc}{{\utwi{c}}}

\newcommand{\bs}{{\utwi{s}}}

\newcommand{\bw}{{\utwi{w}}}
\newcommand{\bx}{{\utwi{x}}}
\newcommand{\by}{{\utwi{y}}}

\begin{document}
     \title{\bf Developing optimal nonlinear scoring
function for protein design }

      \author{\bf Changyu Hu, Xiang Li and Jie Liang
\thanks{Corresponding author.  Phone:
      (312)355--1789, fax: (312)996--5921, email: {\tt
      jliang@uic.edu}} \\ Department of Bioengineering, SEO, MC-063 \\
      University of Illinois at Chicago\\ 851 S.\ Morgan Street, Room
      218 \\ Chicago, IL 60607--7052, U.S.A.\\
Accepted by {\it Bioinformatics}.
}  \date{
      \today}

      \maketitle

\abstract{ \sf
\noindent 
{\bf Motivation.}  Protein design aims to identify sequences
compatible with a given protein fold but incompatible to any
alternative folds.  To select the correct sequences and to guide the
search process, a design scoring function is critically important.
Such a scoring function should be able to characterize the global
fitness landscape of many proteins simultaneously.
\vspace*{.1in}

\noindent {\bf Results.}  To find optimal design scoring functions, we
introduce two geometric views and propose a formulation using mixture
of nonlinear Gaussian kernel functions.  We aim to solve a simplified
protein sequence design problem. Our goal is to distinguish each
native sequence for a major portion of representative protein
structures from a large number of alternative decoy sequences, each a
fragment from proteins of different fold. Our scoring function
discriminate perfectly a set of 440 native proteins from 14 million
sequence decoys. We show that no linear scoring function can succeed
in this task.  In a blind test of unrelated proteins, our scoring
function misclassfies only 13 native proteins out of 194.  This
compares favorably with about $3-4$ times more misclassifications
when optimal linear functions reported in literature are used.  We
also discuss how to develop protein folding scoring function.
}
\vspace*{.1in}


\section{Introduction}
The problem of protein sequence design aims to identify sequences
compatible with a given protein fold and incompatible to alternative
folds \citep{Drexler81_PNAS,Pabo83_Nature,DeGrado99_ARB}.  It is also
called the inverse protein folding problem.  This is a fundamental
problem and has attracted considerable interest
\citep{YueDill92_PNAS,Shakh98_FD,Li96_Science,Deutsch96_PRL,KoehlLevitt99_JMB,KoehlLevitt99b_JMB}.
The ultimate goal of protein design is to engineer protein molecules
with improved activities or with acquired new functions.  There have
been many importantdesign studies, including the design
of novel hydrophobic core \citep{DesjarlaisHandel95_PS,Lazar97_PS}, the
design and experimental validation of an entire protein for specified
backbone \citep{DahiyatMayo97_Science}, the design of a novel alpha helical
protein \citep{Emberly02_PNAS}, the design and validation of a
protein adopting a completely new fold unseen in nature
\citep{Kuhlman03_Science}, and a soluble analog of membrane potassium
channel \citep{Slovic04_PNAS}.

A successful protein design strategy needs to solve two problems.
First, it needs to explore both the sequence and structure search
space and efficiently generates candidate sequences.  Second, a
scoring function or fitness function needs to identify sequences that
are compatible with the desired template fold (the ``design in''
principle) but are incompatible with any other competing folds (the
``design out'' principle)
\citep{YueDill92_PNAS,KoehlLevitt99_JMB,KoehlLevitt99b_JMB}.  To
achieve this, an ideal scoring function would maximize the
probabilities of protein sequences taking their native fold, and
reduce the probability that these sequences take any other fold.
Because many protein sequences with low sequence identity can adopt
the same protein fold, a full-fledged design scoring function should
identify all sequences that fold into the same desired structural fold
from a vast number of sequences that do fold into alternative
structures, or that do not fold.

Several design scoring functions have been developed based on physical
models .  For redesigning protein cores, hydrophoicity and packing
specificity are the main ingredients of the scoring functions
\citep{DesjarlaisHandel95_PS}.  Van der Waals interactions and
electrostatics have also been incorporated for protein design
\citep{KoehlLevitt99_JMB,KoehlLevitt99b_JMB}.  A combination of terms
including Lennard-Jones potential, repulsion, Lazaridis-Karplus
implicit solvation, approximated electrostatic interactions, and
hydrogen bonds is used in an insightful computational protein design
experiment \cite{Kuhlman00_PNAS}.  Models of solvation energy based on
surface area is a key component of several other design scoring
functions \citep{Wernisch00_JMB,KoehlLevitt99_JMB,KoehlLevitt99b_JMB}.

A variety of empirical scoring functions based on known protein
structures have also been developed for coarse-grained models of
proteins.  In this case, proteins are not represented in atomic
details but are represented at residue level.  Because of the
coarse-grained nature of the protein representation, these scoring
functions allow rapid exploration of the search space of the main
factors important for proteins, and can provide good initial solutions
for further refinement where models with atomistic details can be
used.  

Many empirical scoring functions were originally developed for the
purposes of protein folding and structure prediction. Because the
principles are very similar, they are often used directly for protein
design.  One prominent class of  empirical scoring functions are
knowledge-based scoring functions, which are derived from statistical analysis of
database of protein structures
\citep{TanakarScheraga76,Miyazawa85_M,SamudralaMoult98_JMB,LuSkolnick01_Proteins}.
Here the interactions between a pair of
residues are estimated from its relative frequency in database when
compared with a reference state or a null model.  This approach has
found many successfully applications
\citep{Miyazawa96_JMB,SamudralaMoult98_JMB,LuSkolnick01_Proteins,Wodak93_COSB,Sippl95_COSB,Lerner95_Proteins,Jernigan96_COSB,Simons99_Proteins,LiHuLiang03_Proteins}.
However, there are several conceptual difficulties with this
approach. These include the neglect of chain connectivity in the
reference state, and the problematic implicit assumption of Boltzmann
distribution \citep{ThomasDill96_JMB,ThomasDill96_PNAS,Ben-Naim97}.  

An alternative approach for empirical scoring function is to find a
set of parameters such that the scoring functions are optimized by
some criterion, {\it e.g.}, maximized score difference between native
conformation and a set of alternative (or decoy) conformations
\citep{Goldstein92_PNAS,MaiorovCrippen92_JMB,ThomasDill96_PNAS,TobiElber00_Proteins_1,Vendruscolo98_JCP,Vendruscolo00_Proteins,Bastolla01_Proteins,Dima00_PS,Micheletti01_Proteins}.
This approach has been shown to be effective in fold recognition,
where native structures can be identified from alternative
conformations \citep{Micheletti01_Proteins}.  However, if a large
number of native protein structures are to be simultaneously
discriminated against a large number of decoy conformations, no such
scoring functions can be found
\citep{Vendruscolo00_Proteins,TobiElber00_Proteins_1}. Similar
conclusion is found in the present study for protein design, where we
find that no linear design scoring function can simultaneously
discriminate a large number of native proteins from seqeunce decoys.
A recent criticism is that it is impossible to predict stability
changes due to mutation using contact-based scoring function
\cite{Khatun04-JMB}.

There are three key steps in developing effective empirical scoring
function using optimization: (1) the functional form, (2) the
generation of a large set of decoys for discrimination, and (3) the
optimization techniques.  The initial step of choosing an appropriate
functional form is often straightforward.  Empirical pairwise scoring
functions are usually all in the form of weighted linear sum of
interacting residue pairs (see reference \citep{Fain02_PS} for an
exception).  In this functional form, the weight coefficients are the
parameters of the scoring function, which are optimized for
discrimination.  The same functional form is also used in statistical
potential, where the weight coefficients are derived from database
statistics.  The optimization techniques that have been used include
perceptron learning and linear programming
\citep{TobiElber00_Proteins_1,Vendruscolo00_Proteins}.  The objectives
of optimization are often maximization of score gap between native
protein and the average of decoys, or score gap between native and
decoys with lowest score, or the $z$-score of the native protein
\citep{Goldstein92_PNAS,Koretke96,Koretke98_PNAS,Hao96_PNAS,MirnyShkh96_JMB}.

In this work, we study a simplified version of the protein design
problem.  Our goal is to develop a globally applicable scoring
function for characterizng the fintness landscape of many proteins
simultaneously.  Specifically, we aim to identify a protein sequence
that is compatible with a given three-dimensional coarse-grained
structure from a set of protein sequences that are taken from protein
structures of different folds.  In Conclusion, we discuss how to proceed to
develop a full-fledged fitness function that discriminate similar and
dissimilar sequences adopting the same fold against all sequences that
adopt different folds and sequences that do not fold ({\it e.g.}, all
hydrophobes).  In this study, we do not address the problem of how to
generate candidate template fold or candidate sequence by searching
either the conformation space or the sequence space.

To develop empirical scoring function that improves discrimination of
native protein sequence, we explore in this study an alternative
formulation of protein scoring function, in the form of mixture of
nonlinear Gaussian kernel functions.  We also use a different
optimization technique based on quadratic programming.  Instead of
maximizing the score gap, here an objective function related to bounds of
expected classification errors is optimized
\citep{VapChe74,Vapnik95,Burges98,ScholkopfSmola02}.

Experimentation with the nonlinear function developed in this study
shows that it can discriminate simultaneous 440 native proteins
against 14 million sequence decoys.  In contrast, we cannot obtain a
perfect weighted linear sum scoring function using the
state-of-the-art interior point solver of linear programming following
\citep{TobiElber00_Proteins_1,Meller02_JCC}.  We also perform blind
tests for native sequence recognition.  Taking 194 proteins unrelated
to the 440 training set proteins, the nonlinear scoring function
achieves a success rate of 93.3\% in sequence design.  This result
compares favorably with optimal linear scoring function (80.9\% and
73.7\% success rate) and statistical potential (58.2\%)
\citep{TobiElber00_Proteins_1,Bastolla01_Proteins,Miyazawa96_JMB}.

The rest of the paper is organized as follows.  We first describe
theory and model of linear and nonlinear function, including the
kernel model and the optimization technique.  We then explain details
of computation.  We further describe experimental results of learning
and results of blind test.  We conclude with discussion about how
these ides may be applicable for developing protein folding scoring
function.

\section{Theory and Models}

\paragraph{Modeling Protein Design Scoring Function.}
To model protein computationally, we first need a method to describe
its geometric shape and its sequence of amino acid residues.
Frequently, a protein is represented by a $d$-dimensional vector $\bc
\in \real^d$.  For example, a method that is widely used is to count
nonbonded contacts of various types of amino acid residue pairs in a
protein structure.  In this case, the count vector $\bc \in \real^d,
d=210$, is used as the protein descriptor.  Once the structural
conformation of a protein $\bs$ and its amino acid sequence $\ba$ is
given, the protein description $f: (\bs, \ba) \mapsto \real^d$ will
fully determine the $d$-dimensional vector $\bc$.  In the case of
contact vector, $f$ corresponds to the mapping provided by specific
contact definition, {\it e.g.}, two residues are in contact if their
distance is below a specific cut-off threshold distance.

To develop scoring functions for our simplified problem, namely, a
scoring function that allows the search and identification of
sequences most compatible with a specific given coarse-grain
three-dimensional structure, we use a model analogous to the Anfinsen
experiments in protein folding.  We require that the native amino acid
sequence $\ba_N$ mounted on the native structure $\bs_N$ has the
best (lowest) fitness score compared to a set of alternative sequences
(sequence decoys) taken from unrelated proteins known to fold into a
different fold ${\cal D} = \{\bs_N, \ba_D\}$ when mounted on the same
native protein structure $\bs_N$:
\[
H(f(\bs_N, \ba_N)) < H(f(\bs_N, \ba_D)) \quad \mbox{for all } \ba_D \in {\cal D}.
\]
Equivalently, the native sequence will have the highest probability
to fit into the specified native structure.  This is the same
principle described in
\citep{ShakhGutin93_PNAS,Deutsch96_PRL,Li96_Science}.  
Sometimes we can further require that the score difference must be
greater than a constant $b>0$:
\[
H(f(\bs_N, \ba_N)) + b < H(f(\bs_N, \ba_D)) \quad \mbox{for all }
(\bs_D, \ba_N) \in {\cal D}.
\]

A widely used functional form for protein scoring function $H$ is the
weighted linear sum of pairwise contacts
\citep{TanakarScheraga76,Miyazawa85_M,TobiElber00_Proteins_1,Vendruscolo98_JCP,SamudralaMoult98_JMB,LuSkolnick01_Proteins}.
The linear sum score $H$ is:
\begin{equation}
H(f(\bs, \ba)) = H(\bc) = \bw \cdot \bc,
\label{linear}
\end{equation}
where ``$\cdot$'' denotes inner product of vectors.  As soon as the
weight vector $\bw$ is specified, the scoring function is fully
defined.  Much work has been done using this class of design function
of linear sum of contact pairs \citep{ShakhGutin93_PNAS,Deutsch96_PRL}.
For such linear scoring functions, the basic requirement for design
scoring function is then:
\[
\bw \cdot (\bc_N - \bc_D) < 0,
\]
or 
\begin{equation}
\bw \cdot (\bc_N - \bc_D) + b < 0,
\label{eq:gap}
\end{equation}
if we require that the score difference between a native protein and a
decoy must be greater than a real value $b$.  The goal here is to
obtain a scoring function to discriminate native proteins from
decoys. An ideal scoring function therefore would assign the value
``$-1$'' for native structure/sequence, and the value ``$+1$'' for
decoys.

\begin{figure}[t!]
\centerline{\epsfig{figure=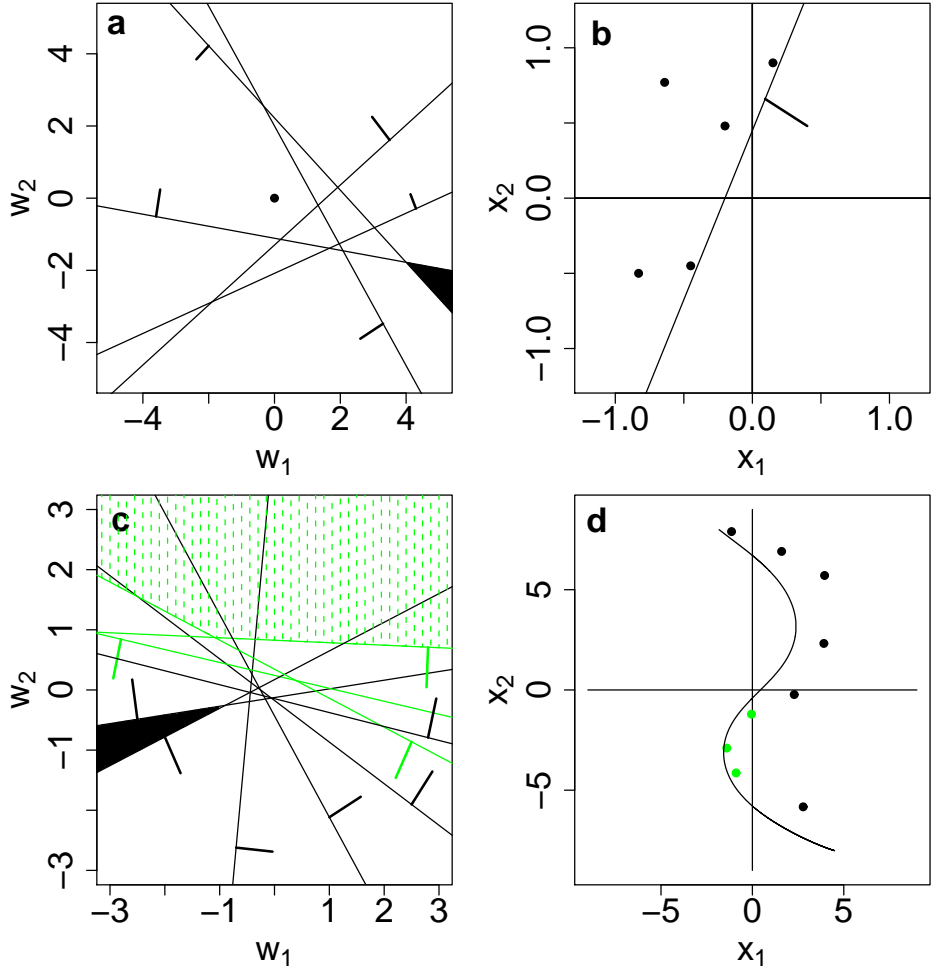,width=5in}}
\caption{\small \sf
Geometric views of the inequality
requirement for protein scoring function.  Here we use a
two-dimensional toy example for illustration.  (a). In the first
geometric view, the space $\real^2$ of $\bw = (w_1, w_2)$ is divided
into two half-spaces by an inequality requirement, represented as a
hyperplane $\bw \cdot (\bc_N - \bc_D) + b < 0$. The hyperplane, which
is a line in $\real^2$, is defined by the normal vector $(\bc_N -
\bc_D)$, and its distance $b/||\bc_N - \bc_D ||$ from the origin.  In
this figure, this distance is set to 1.0.  The normal vector is
represented by a short line segment whose direction points away from
the straight line.  A feasible weight vector $\bw$ is located in the
half-space opposite to the direction of the normal vector $(\bc_N -
\bc_D)$.  With the given set of inequalities represented by the lines,
any weight vector $\bw$ located in the shaped polygon can satisfy all
inequality requirement and provides a linear scoring function that
has perfect discrimination.  (b). A second geometric view of the
inequality requirement for linear protein scoring function.  The space
$\real^2$ of $\bx=(x_1, x_2)$, where $\bx \equiv (\bc_N - \bc_D)$, is
divided into two half-spaces by the hyperplane $\bw \cdot (\bc_N -
\bc_D) + b < 0$. Here the hyperplane is defined by the normal vector
$\bw$ and its distance $b/||\bw ||$ from the origin.  All points
$\{\bc_N - \bc_D\}$ are located on one side of the hyperplane away
from the origin, therefore satisfying the inequality requirement.
That is, a linear scoring function $\bw$ such as the one represented
by the straight line in this figure can have perfect discrimination.
(c). In the second toy problem, a set of inequalities are represented
by a set of straight lines according to the first geometric view.  A
subset of the inequalities require that the weight vector $\bw$ to be
located in the shaded convex polygon on the left, but another subset
of inequalities require that $\bw$ to be located in the dashed convex
polygon on the top.  Since these two polygons do not intersect,
there is no weight vector $\bw$ that can satisfy all inequality
requirements.  That is, no linear scoring function can classify these
decoys from native protein.  (d).  According to the second geometric
view, no hyperplane can separate all points $\{\bc_N - \bc_D\}$ from
the origin.  But a nonlinear curve formed by a mixture of Gaussian
kernels can have perfect separation of all vectors $\{ \bc_N -
\bc_D\}$ from the origin: It has perfect discrimination.
}
%\vspace*{.1in}
\label{Fig:geom1}
\end{figure}

\paragraph{Two Geometric Views of Linear Protein Folding Potentials.}

There is a natural geometric view of the inequality requirement for
weighted linear sum scoring functions.  A useful observation is that
each of the inequalities divides the space of $\real^d$ into two halfs
separated by a hyperplane (Fig~\ref{Fig:geom1}a). The hyperplane for
Equation (\ref{eq:gap}) is defined by the normal vector $(\bc_N -
\bc_D)$ and its distance $b/||\bc_N - \bc_D ||$ from the origin.  The
weight vector $\bw$ must be located in the half-space opposite to the
direction of the normal vector $(\bc_N - \bc_D)$.  This half-space can
be written as $\bw \cdot (\bc_N - \bc_D) + b < 0$.  When there are
many inequalities to be satisfied simultaneously, the intersection of
the half-spaces forms a convex polyhedron \citep{Edels87}. If the
weight vector is located in the polyhedron, all the inequalities are
satisfied.  Scoring functions with such weight vector $\bw$ can
discriminate the native protein sequence from the set of all decoys.
This is illustrated in Fig~\ref{Fig:geom1}a for a two-dimensional toy
example, where each straight line represents an inequality $\bw \cdot
(\bc_N - \bc_D) + b < 0$ that the scoring function must satisfy.  

For each native protein $i$, there is one convex polyhedron ${\cal
P}_i$ formed by the set of inequalities associated with its decoys.
If a scoring function can discriminate simultaneously $n$ native
proteins from a union of sets of sequence decoys, the weight vector
$\bw$ must be located in a smaller convex polyhedron $\cal P$ that is
the intersection of the $n$ convex polyhedra:
\[
\bw \in {\cal P} = \bigcap_{i=1}^n
{\cal P}_i.
\]

There is yet another geometric view of the same inequality
requirements.  If we now regard $(\bc_N - \bc_D)$ as a point in
$\real^d$, the relationship $\bw \cdot (\bc_N - \bc_D) + b < 0$ for
all sequence decoys and native proteins requires that all points
$\{\bc_N - \bc_D\}$ are located on one side of a different hyperplane,
which is defined by its normal vector $\bw$ and its distance
$b/||\bw||$ to the origin (Fig~\ref{Fig:geom1}b).  We can show that
such a hyperplane exists if the origin is not contained within the
convex hull of the set of points $\{ \bc_N - \bc_D\}$ (see Appendix).

The second geometric view looks very different from the first view.
However, the second view is dual and mathematically equivalent to the
first geometric view.  In the first view, a point $\bc_N - \bc_D$
determined by the structure-decoy pair $c_N = (\bs_N,\ba_N)$ and
$c_D=(\bs_N, \ba_D)$ corresponds to a hyperplane representing an
inequality, a solution weight vector $\bw$ corresponds to a point
located in the final convex polyhedron.  In the second view, each
structure-decoy pair is represented as a point $ \bc_N -\bc_D$ in
$\real^d$, and the solution weight vector $\bw$ is represented by a
hyperplane separating all the points ${\cal C} = \{\bc_N - \bc_D \}$
from the origin.

\paragraph{Optimal Linear Scoring Function.}
Several optimization methods have been applied to find the weight
vector $\bw$ of linear scoring function.  The Rosenblantt perceptron
method works by iteratively updating an initial weight vector $\bw_0$
\citep{Vendruscolo98_JCP,Micheletti01_Proteins}.  Starting with a
random vector, {\it e.g.}, $\bw_0 ={\bf 0}$, one tests each native
protein and its decoy structure.  Whenever the relationship $ \bw
\cdot (\bc_N - \bc_D) + b < 0$ is violated, one updates $\bw$ by
adding to it a scaled violating vector $\eta \cdot (\bc_N - \bc_D)$.
The final weight vector is therefore a linear combination of protein
and decoy count vectors:
\begin{equation}
\bw = \sum \eta (\bc_N - \bc_D) = \sum_{N \in {\cal N}} \alpha_N
\bc_N - \sum_{D \in {\cal D}} \alpha_D  \bc_D.
\label{LinearEq}
\end{equation}
Here $\cal N$ is the set of native proteins, and $\cal D$ is the set
of decoys.  The set of coefficients $\{\alpha_N \} \cup \{\alpha_D\}$
gives a dual form representation of the weight vector $\bw$, which is
an expansion of the training examples including both native and decoy
structures.

According to the first geometric view, if the final convex polyhedron $\cal
P$ is non-empty, there can be infinite number of choices of $\bw$, all
with perfect discrimination.  But how do we find a weight vector $\bw$
that is optimal?  This depends on the criterion for optimality.  For
example, one can choose the weight vector $\bw$ that minimizes the
variance of score gaps between decoys and natives: $ \arg_\bw \min
\frac{1}{|{\cal D}|} \sum \left( \bw \cdot (c_N - c_D)\right)^2 -
\left[ \frac{1}{|{\cal D}|}\sum_D \left(\bw \cdot (\bc_N
-\bc_D)\right) \right]^2 $ as used in reference
\citep{TobiElber00_Proteins_1}, or minimizing the $Z$-score of a large
set of native proteins, or minimizing the $Z$-score of the native
protein and an ensemble of decoys
\citep{ChiuGoldstein98_FD,MirnyShkh96_JMB}, or maximizing the ratio $R$
between the width of the distribution of the score and the average
score difference between the native state and the unfolded ones
\citep{Goldstein92_PNAS,Hao99}.
A series of important works using perceptron learning and other
optimization techniques
\citep{FriedrichsWolynes89_Science,Goldstein92_PNAS,TobiElber00_Proteins_1,Vendruscolo98_JCP,Dima00_PS}
showed that effective linear sum scoring functions can be obtained.

Here we describe yet another optimality criterion according to the
second geometric view.  We can choose the hyperplane $(\bw, b)$ that
separates the points $\{\bc_N - \bc_D\}$ with the largest distance to
the origin.  Intuitively, we want to characterize proteins with a
region defined by the training set points $\{\bc_N - \bc_D \}$.  It is
desirable to define this region such that a new unseen point drawn
from the same protein distribution as $\{\bc_N - \bc_D \}$ will have a
high probability to fall within the defined region. Non-protein
points following a different distribution, which is assumed to be
centered around the origin when no {\it a priori\/} information is
available, will have a high probability to fall outside the defined
region.  In this case, we are more interested in modeling the region
or support of the distribution of protein data, rather than estimating
its density distribution function.  For linear scoring function, regions are
half-spaces defined by hyperplanes, and the optimal hyperplane $(\bw,
b)$ is then the one with maximal distance to the origin.  This is related
to the novelty detection problem and single-class support vector
machine studied in statistical learning theory
\citep{VapChe64,VapChe74,ScholkopfSmola02}.  In our case, any
non-protein points will need to be detected as outliers from the
protein distribution characterized by $\{\bc_N - \bc_D \}$.  Among all
linear functions derived from the same set of native proteins and
decoys,  an optimal weight vector $\bw$ is likely to have the least
amount of mislabellings.
The optimal weight vector $\bw$ can be found by solving the following
quadratic programming problem:
\begin{eqnarray}
\mbox{Minimize } & \frac{1}{2} || \bw||^2
\\
\mbox{subject to} & \bw \cdot (\bc_N - \bc_D) + b < 0 \mbox{ for all }
N \in {\cal N} \mbox{ and } D \in {\cal D}.
\label{Eqn:PrimalLinear}
\end{eqnarray}
The solution maximizes the distance $b/||\bw||$ of the plane $(\bw,
b)$ to the origin.  
We obtained the solution
by solving the following support vector machine problem:
\begin{equation}
\begin{array}{ll}
\mbox{Minimize} & \frac{1}{2}\|\bw\|^2 \\
\mbox{subject to} &\bw \cdot \bc_N + d \le -1 \\
&\bw \cdot \bc_D +d \ge 1,
\end{array}
\label{Eq:svm}
\end{equation}
where $d>0$. Note that a solution of Problem (\ref{Eq:svm}) satisfies
the constraints in Inequalities (\ref{Eqn:PrimalLinear}), since
subtracting the second inequality here from the first inequality in
the constraint conditions of (\ref{Eq:svm}) will give us $\bw \cdot
(\bc_N - \bc_D) +2 \le 0$.

\paragraph{Nonlinear Scoring Function.}
However, it is possible that the weight vector $\bw$ does not exist,
{\it i.e.}, the final convex polyhedron ${\cal P} = \bigcap_{i=1}^n
{\cal P}_i$ may be an empty set.  First, for a specific native protein
$i$, there may be severe restriction from some inequality constraints,
which makes ${\cal P}_i$ an empty set.  Some decoys are very difficult
to discriminate due to perhaps deficiency in protein representation.
In these cases, it is impossible to adjust the weight vector so the
native protein has a lower score than the sequence decoy.
Figure~\ref{Fig:geom1}c shows a set of inequalities represented by
straight lines according to the first geometric view.  A subset of
inequalities (black lines) require that the weight vector $\bw$ to be
located in the shaded convex polygon on the left, but another subset
of inequalities (green lines) require that $\bw$ to be located in the
dashed convex polygon on the top.  Since these two polygons do not
intersect, there is no weight vector that can satisfy all these
inequality requirements.  That is, no linear scoring function can
classify all decoys from native protein.  According to the
second geometric view (Figure~\ref{Fig:geom1}d), no hyperplane can
separate all points (black and green) $\{\bc_N - \bc_D\}$ from the
origin.

Second, even if a weight vector $\bw$ can be found for each native
protein, {\it i.e.}, $\bw$ is contained in a nonempty polyhedron, it
is still possible that the intersection of $n$ polyhedra is an empty
set, {\it i.e.}, no weight vector can be found that can discriminate
all native proteins against the decoys simultaneously.
Computationally, the question whether a solution weight vector $\bw$
exists can be answered unambiguously in polynomial time
\citep{Karmarkar84}, and results described later in this study show
that when the number of decoys reaches millions, no such weight vector
can be found.

A fundamental reason for this failure is that the functional form of
linear sum is too simplistic.  It has been suggested that additional
decriptors of protein structures such as higher order interactions
({\it e.g.},  three-body or four-body contacts) should be incorporated in
protein description \citep{Betancourt99_PS,Munson97,Zheng97}. Functions
with polynomial terms using upto 6 degree of Chebyshev expansion has
also been used to represent pairwise interactions in protein folding
\citep{Fain02_PS}.

Here we propose an alternative approach.  In this study we still limit
ourselves to pairwise contact interactions, although it can be
naturally extended to include three or four body interactions
\citep{LiLiang04-triple}.  We introduce a nonlinear scoring 
function analogous to the dual form of the linear function in Equation
(\ref{LinearEq}), which takes the following form:
\begin{equation}
H(f(\bs, \ba)) =  H(\bc) =
\sum_{D \in {\cal D}}\alpha_D  K(\bc, \bc_D) -
\sum_{N \in {\cal N}}\alpha_N  K(\bc, \bc_N),
\label{nonlinear}
\end{equation}
where $\alpha_D \ge 0$ and $\alpha_N \ge 0$ are parameters of the
scoring function to be determined, and $\bc_D = f(\bs_N, \ba_D)$ from
the set of decoys $\cal D = \{ (\bs_N, \ba_D )\}$ is the contact
vector of a sequence decoy $D$ mounted on a native protein structure
$\bs_N$, and $\bc_N = f(\bs_N, \ba_N)$ from the set of native training
proteins ${\cal N} = \{(\bs_N, \ba_N) \}$ is the contact vector of a
native sequence $\ba_N$ mounted on its native structure $\bs_N$.  In
this study, all decoy sequence $\{ \ba_D\}$ are taken from real
proteins possessing different fold structures.  The difference of this
functional form from linear function in Equation (\ref{LinearEq}) is
that a kernel function $K(\bx, \by)$ replaces the linear term.  A
convenient kernel function $K$ is:
\[K(\bx, \by) =
e^{-||\bx -\by||^2/2\sigma^2} \quad \mbox{for any vectors $\bx$ and $\by \in {\cal N} \bigcup {\cal D}$},
\]
where $\sigma^2$ is a constant.  Intuitively, the surface of the
scoring function has smooth Gaussian hills of height $\alpha_D$
centered on the location $\bc_D$ of decoy protein $D$, and has
smooth Gaussian cones of depth $\alpha_N$ centered on the location
$\bc_N$ of native structures $N$. Ideally, the value of the scoring
function will be $-1$ for contact vectors $\bc_N$ of native proteins,
and will be $+1$ for contact vectors $\bc_D$ of decoys.

\paragraph{Optimal Nonlinear Scoring Function.}

To obtain the nonlinear scoring function, our goal is to find a set of
parameters $\{\alpha_D, \alpha_N\}$ such that $ H(f(\bs_N, \ba_N))$
has value close to $-1$ for native proteins, and the decoys have
values close to $+1$.  There are many different choices of
$\{\alpha_D, \alpha_N \}$.  We use an optimality criterion originally
developed in statistical learning theory
\citep{Vapnik95,Burges98,ScholkopfSmola02}.  First, we note that we
have implicitly mapped each structure and decoy from $\real^{210}$
through the kernel function of $K(\bx, \by) = e^{-||\bx
-\by||^2/2\sigma^2}$ to another space with dimension as high as tens
of millions.  Second, we then find the hyperplane of the largest
margin distance separating proteins and decoys in the space
transformed by the nonlinear kernel.  That is, we search for a
hyperplane with equal and maximal distance to the closest native
proteins and the closest decoys in the transformed high dimensional
space.  Such a hyperplane can be found by obtaining the parameters $\{
\alpha_D \}$ and $\{ \alpha_N \}$ from solving the following Lagrange
dual form of quadratic programming problem:
\begin{eqnarray*}
\mbox{Maximize } & \sum_{
\substack{
i\in  {\cal N} \cup {\cal D},
}
}  
\alpha_i -
 \frac{1}{2}\sum_{
\substack{
i,j\in {{\cal N}\cup {\cal D}}
}
}
y_i y_j \alpha_i \alpha_j  e^{-||\bc_i - \bc_j||^2/2\sigma^2}
\\
\mbox{subject to} & 0\le \alpha_i \le C,\\
\end{eqnarray*}
where $C$ is a regularizing constant that limits the influence of each
misclassified protein or decoy
\citep{VapChe64,VapChe74,Vapnik95,Burges98,ScholkopfSmola02}, and $y_i
=-1$ if $i$ is a native protein, and $y_i= +1$ if $i$ is a decoy.
These parameters lead to optimal discrimination of an unseen test set
\citep{VapChe64,VapChe74,Vapnik95,Burges98,ScholkopfSmola02}.  When
projected back to the space of $\real^{210}$, this hyperplane becomes
a nonlinear surface.  For the toy problem of Figure~\ref{Fig:geom1},
Figure~\ref{Fig:geom1}d shows that such a hyperplane becomes a
nonlinear curve in $\real^2$ formed by a mixture of Gaussian kernels.
It separates perfectly all vectors $\{ \bc_N - \bc_D\}$ (black and
green) from the origin.  That is, a nonlinear scoring function can
have perfect discrimination.

\section{Computational Methods}
\paragraph{Alpha Contact Maps.}
Because protein molecules are formed by thousands of atoms, their
shapes are complex.  In this study we use the count vector of pairwise
contact interactions after normalization by the chain length of the
protein \citep{Edels95_DCG,Liang98a_Proteins}.  Here contacts are
derived from the edge simplices of the alpha shape of a protein
structure \citep{LiHuLiang03_Proteins}.  These edge simplices represent
nearest neighbor interactions that are in physical contacts.  They
encode precisely the same contact information as a subset of the edges
in the Voronoi diagram of the protein molecule.  These Voronoi edges
are shared by two interacting atoms from different residues, but
intersect with the body of the molecule modeled as the union of atom
balls.  Statistical potential based on edge simplices has  been
developed \citep{LiHuLiang03_Proteins}.  We refer to references
\citep{Edels95_DCG,Liang98a_Proteins} for further theoretical and
computational details.

\paragraph{Generating Sequence Decoys by Threading.}
Maiorov and Crippen introduced the gapless threading method to
generate a large number of decoys \citep{MaiorovCrippen92_JMB}.  The
sequence of a smaller protein $\ba_N$ is threaded through the
structure of an unrelated larger protein and takes the conformation
$\bs_D$ of a fragment with the same length from the larger protein
\citep{MaiorovCrippen92_JMB}.  Along the way, the sequence of the
smaller protein can take the conformations of many fragments of the
larger protein, each becomes a structure decoy.

\begin{figure}[t!]
      \centerline{\epsfig{figure=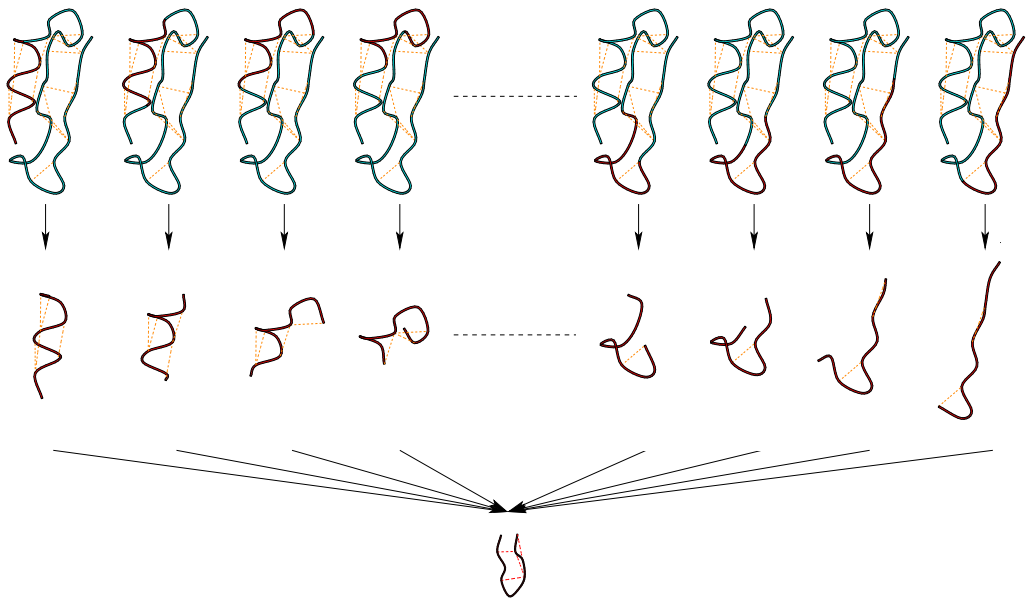,width=5.5in}}
\caption{\sf
Decoy generation by gapless
threading.  Sequence decoys can be generated by threading the sequence
of a larger protein to the structure of an unrelated smaller protein.
}
      \label{Fig:Thread.structure}
\end{figure}

We can generate sequence decoys in an analogous way, as already
suggested in \citep{Jones92_Nature,Munson97}.  We thread the sequence
of a larger protein through the structure of a smaller protein, and
obtain sequence decoys by mounting a fragment of the sequence of the
large protein to the full structure of the small protein.  We
therefore have for each native protein $(\bs_N, \ba_N)$ a set of
sequence decoys $(\bs_N, \ba_D)$ (Fig~\ref{Fig:Thread.structure}).
Because all native contacts are retained in this case,
sequence decoys obtained by gapless threading are far more challenging
than structure decoys generated by gapless threading.

\paragraph{Protein Data.}
Following reference \citep{Vendruscolo00_Proteins.pair}, we use protein structures
contained in the {\sc Whatif} database  \citep{WHATIF} in this study.
{\sc Whatif} database contains a representative set of sequence-unique
protein structures generated from X-ray crystallography.  Structures
selected for this study all have pairwise sequence identity $< 30\%$,
R-factor $<0.21$, and resolution $<2.1$\AA.  {\sc Whatif} database
contains less structures than {\sc Pdbselect} because the R-factor and
resolution criteria are more stringent \citep{WHATIF}.  Nevertheless,
it provides a good representative set of currently all known protein
structures.

We use a list of 456 proteins kindly provided by Dr.\ Vendruscolo,
which was compiled from the 1998 release ({\sc Whatif98}) of the {\sc
Whatif} database \citep{Vendruscolo00_Proteins}.  There are 192
proteins with multiple chains in this dataset.  Some of them have
extensive interchain contacts.  For these proteins, it is possible
that their conformations may be different if there are no interchain
contacts present.  We use the criterion of {\it Contact Ratio} to
remove proteins that have extensive interchain contacts.  Contact
Ratio is defined here as the number of interchain contacts divided by
the total number of contacts a chain makes. For example, protein {\tt
1ept} has four chains A,B,C, and D. The intra chain contact number of
chain B is 397.  Contacts between chain A and chain B is 178, between
B and C is 220, between B and other heteroatoms is 11.  The Contact
Ratio of chain B is therefore $ (178+220+11)/ (397+178+220+11) = 51 \%
$.  Thirteen protein chains are removed because they all have Contact
Ratio $ > 30 \% $.  We further remove three proteins because each has
$>10\% $ of residues missing with no coordinates in the Protein Data
Bank file.  The remaining set of 440 proteins are then used as
training set for developing both folding and design scoring functions.
Using threading method described earlier, we generated a set of
14,080,766 sequence decoys.

\paragraph{Learning Linear Scoring Function.}
For comparison, we have also developed optimal linear scoring function following
the method and computational procedure described in reference
\citep{TobiElber00_Proteins_1}. We apply the interior point method as
implemented in BPMD package by M\'esz\'aros \citep{Meszaros96_CMA} to search
for a weight vector $\bw$.  We use two different optimization criteria
as described in reference \citep{TobiElber00_Proteins_1}.  The first
is:
\begin{eqnarray*}
\mbox{Identify} & \bw 
\\
\mbox{subject to} & \bw \cdot (\bc_N - \bc_D) < \epsilon  \quad \mbox{ and } |w_i| \le 10,\\
\end{eqnarray*}
where $w_i$ denotes the $i$-th component of weight vector $\bw$, and
$\epsilon = 1 \times 10^{-6}$.  Let $ {\cal C} = \{ c_N - c_D \}$, and
$|\cal C|$ the number of decoys.  The second optimization criterion
is:
\begin{eqnarray*}
\mbox{Minimize }  & 
 \min \frac{1}{|{\cal C}|} \sum \left( \bw \cdot (c_N - c_D)\right)^2 - 
                         \left[ \frac{1}{|{\cal C}|}\sum  \left(\bw \cdot (\bc_N -\bc_D)\right) \right]^2 
 \\
\mbox{subject to} & \bw \cdot (\bc_N - \bc_D) < \epsilon. \\
\end{eqnarray*}

\paragraph{Learning Nonlinear Kernel Scoring Function.}
We use {\sc SVMlight} ({\tt http://svmlight.joachims.org/})
\citep{Joachims99} with Gaussian kernels and a training set of 440
native proteins plus 14,080,766 decoys to obtain the optimized
parameter $\{\alpha_N, \alpha_D\}$.
The regularization constant $C$  takes  default value,  which is
estimated from the training set ${{\cal N} \cup {\cal D}}$:
\begin{equation} 
C =  |{\cal N} \cup {\cal D}|^2 / \left [ \sum_{\bx \in {\cal N} \cup {\cal D}} \sqrt{K(\bx,
\bx)-2 \cdot K(\bx,{\bf 0}) + K({\bf 0, 0})} \right ]
^2.
\end{equation}

Since we cannot load all 14 millions decoys into computer memory
simultaneously, we use a heuristic strategy for training.  Similar to
the procedure reported in \citep{TobiElber00_Proteins_1}, we first
randomly selected a subset of decoys that fits into the computer
memory.  Specifically, we pick every 51st decoy from the list of 14
million decoys.  This leads to an initial training set of 276,095
decoys and 440 native proteins. An initial protein scoring function is
then obtained.  Next the scores for all 14 million decoys and all 440
native proteins are evaluated.  Three decoy sets were collected based
on the evaluation results: the first set of decoys contains the
violating decoys which have lower score than the native structures;
the second set contains decoys with the lowest absolute score, and the
third set contains decoys that participate in $H(\bc)$ as identified
in previous training process.  The union of these three subsets of
decoys are then combined with the 440 native protein as the training
set for the next iteration of learning. This process is repeated until
the score difference to native protein for all decoys are greater than
0.0.  Using this strategy, the number of iterations typically is
between 2 and 10.  During the training process, we set the cost factor
$j$ in {\sc SVMlight} to 120, which is the factor training errors on
native proteins outweighs training errors on decoys.

The value of $\sigma^2$ for the Gaussian kernel $K(\bx, \by) =
e^{-||\bx -\by||^2/2\sigma^2}$ is chosen by experimentation.  If the
value of $\sigma^2$ is too large, no parameter set $\{\alpha_N,
\alpha_D\}$ can be found such that the fitness scoring function can
perfectly classifies the 440 training proteins and their decoys, {\it
i.e.}, the problem is unlearnable. If the value of $\sigma^2$ is too
small, the performance in blind-test will deteriorate.  The final
final design scoring function is obtained with $\sigma^2$ set to
$416.7$.

\section{Results}

\paragraph{Linear Design Scoring Functions.}
To search for the optimal weight vector $\bw$ for design scoring
function, we use linear programming solver based on interior point
method as implemented in BPMD by M\'esz\'aros \citep{Meszaros96_CMA}.
After generating 14,080,766 sequence design decoys for the 440
proteins in the training set, we search for an optimal $\bw$ that can
discriminate native sequences from decoy sequences. That is, we search
for parameters $\bw$ for $H(\bs, \ba) = \bw \cdot \bc$, such that $
\bw \cdot \bc_N < \bw \cdot \bc_D$ for all sequences.  However, we
fail to find a feasible solution for the weight vector $\bw$.  That
is, no $\bw$ exists capable of discriminating perfectly 440
native sequences from the 14 million decoy sequences.  We repeated the
same experiment using a larger set of 572 native proteins from reference
\citep{TobiElber00_Proteins_1} and 28,261,307 sequence decoys.  The
result is also negative.

\paragraph{Nonlinear Kernel Scoring Function.}
To overcome the problems associated with linear function, we use the
set of 440 native proteins and 14 million decoys to derive
nonlinear kernel design functions.  We succeeded in finding a function
in the form of Equation (\ref{nonlinear}) that can discriminate all
440 native proteins from 14 million decoys.

\begin{table}[t!]
\begin{small}
\begin{center}
\caption{\sf
 Details of derivation of nonlinear
 kernel design scoring functions.  The numbers of native proteins and
 decoys with non-zero $\alpha_i$ entering the scoring function are
 listed.  The range of the score values of natives and decoys are also
 listed, as well as the range of the smallest gaps between the scores
 of the native protein and decoy. Details for nonlinear kernel folding
 scoring functon are also listed.}
\label{tab:learning}
\vspace*{.3in}
\begin{tabular}{|c|c|c|c|}
\hline
\multicolumn{2}{|c|}{} &Design Scoring Function & Folding Scoring Function\\ \cline{3-4}
\multicolumn{2}{|c|}{} &  $\sigma^2 = 416.7$ & $\sigma^2 = 227.3$\\ \hline

Num.\ of        & Natives & 220 & 214  \\ \cline{2-4}
Vectors & Decoys  & 1685 & 1362 \\ \hline
Range of        & Natives &  $0.9992 \sim 4.598 $ & $0.9990 \sim  4.215$ \\ \cline{2-4}
Score Values   & Decoys  &  $-9.714 \sim 0.7423$ & $-6.859 \sim 0.3351$\\ \hline
\multicolumn{2}{|c|}{Range of Smallest Score Gap } &  $0.2575 \sim 11.53 $ & $0.8446 \sim 9.816 $ \\
\hline
\end{tabular}
\end{center}
\end{small}
\end{table}

Unlike statistical scoring functions where each native protein in
the database contribute to the empirical scoring function, only a subset of
native proteins contribute and have $\alpha_N \ne 0$.  In addition,
a small fraction of decoys also contribute to the scoring function.
Table~\ref{tab:learning} list the details of the scoring function, including
the numbers of native proteins and decoys that participate in Equation
(\ref{nonlinear}).  These number represent about $ 50\% $ of native
proteins and $<0.1\%$ of decoys from the original training data.

\paragraph{Discrimination Tests for Design Scoring Function.}
Blind test in discriminating native proteins from decoys for an
independent test set is essential to assess the effectiveness of
design scoring functions.  To construct a test set, we first take the
entries in {\sc Whatif99} database that are not present in {\sc
Whatif98}.  After eliminating proteins with chain length less than 46
residues, we obtain a set of 201 proteins.  These proteins all have
$<$ 30\% sequence identities with any other sequence in either the
training set or the test set proteins.  Since 139 of the 201 test
proteins have multiple chains, we use the same criteria applied in
training set selection to exclude 7 proteins with $>30\%$ Contact
Ratio or with $>10\%$ residues missing coordinates in the PDB files.
This leaves a smaller set of test proteins of 194 proteins.  Using
gapless threading, we generate a sets of 3,096,019 sequence decoys
from the set of 201 proteins. This is a superset of the decoy set
generated using 194 proteins.

To test  design scoring functions for discriminating native
proteins from sequence decoys in both the 194 and the 201 test sets, we
take the sequence $\ba$ from the conformation-sequence pair $(\bs_N,
\ba)$ for a protein with the lowest score as the predicted sequence.
If it is not the native sequence $\ba_N$, the discrimination failed
and the design scoring function does not work for this protein.

For comparison, we also test the discrimination results of  optimal
linear scoring function taken as reported in reference
\citep{TobiElber00_Proteins_1}, as well as the statistical potential
developed by Miyazawa and Jernigan.  Here we use the contact
definition reported in \citep{TobiElber00_Proteins_1}, that is, two
residues are declared to be in contact if the geometric centers of
their side chains are within a distance of 2.0 -- 6.4 \AA.

\begin{table}[t!]
\begin{center}
\caption{\sf
The number of misclassified
protein sequences for the test set of 194 proteins and the set of 201
proteins using nonlinear kernel design scoring function, two optimal
linear scoring function taken as reported in
\citep{TobiElber00_Proteins_1}, in Table I of
\citep{Bastolla01_Proteins}, and Miyazawa-Jernigan statistical
potential \citep{Miyazawa96_JMB}.  
The nonlinear kernel
design scoring function has the best performance in blind test and is
the only function that succeeded in perfect discrimination of the 440
native sequences from a set of 14 million sequence decoys.
}
\label{tab:result.design}
\vspace*{.3in}
\begin{tabular}{|c|c|c|}
\hline
    & { Misclassified Natives} & {Misclassified Natives} \\ \hline \hline
{ Kernel Design Scoring Function}  &  13/194 & 19/201 \\ \hline
Tobi \& Elber              &  37/194 & 44/201 \\ \hline
Bastolla {\it et al}            &  51/194 & 54/201 \\ \hline 
Miyazawa \& Jernigan            &  81/194 & 87/201 \\ \hline 
\end{tabular}\\
\end{center}
\end{table}

The nonlinear design scoring function capable of discriminating all of
the 440 native sequences also works well for the test set
(Table~\ref{tab:result.design}).  It succeeded in correctly
identifying 93.3\% (181 out of 194) of native sequences in the
independent test set of 194 proteins.  This compares favorably with
results obtained using optimal linear folding scoring function taken
as reported in \citep{TobiElber00_Proteins_1}, which succeeded in
identifying 80.9\% (157 out of 194) of this test set.  It also has
better performance than optimal linear scoring function based on
calculations using parameters reported in reference
\citep{Bastolla01_Proteins}, which succeeded in identifying 73.7\% (143
out of 194) of proteins in the test set.  The Miyazawa-Jernigan
statistical potential succeeded in identifying 113 native proteins out
of 194) (success rate 58.2\%).

\paragraph{Discirmintating Dissimilar Proteins.}
 
\begin{table}[b!]
\begin{small}
\begin{center}
\caption{
  Discrimination of five large
proteins against (a) design decoys and (b) folding decoys generated by
gapless threading, and against (c) additional design decoys generated
by threading unrelated long proteins (length from 1 124 to 2 459) to
the structures of these five proteins.  Here pdb is the pdb code of
the protein structure, $N$ is the size of protein, $n$ is the number
of decoys, $H$ is the predicted value of the scoring function,
$\Delta_{score}$ is the smallest gap of score between the native
protein and its decoys.  The results show that all decoys can be
discriminated from natives, and the smallest score gaps between native
and decoys are large.
}
\label{tab:largeProteins}
\vspace*{.3in}
\begin{tabular}{|c|r||r|c|c|c|c||r|c|c|}
\multicolumn{10}{c}{  }\\ \hline  
&&&\multicolumn{2}{|c|}{$^a$Design Decoy}&\multicolumn{2}{c||}{$^b$Folding Decoy}&\multicolumn{3}{c|}{$^c$SwissProt Decoy}\\
pdb & $N$ & $n$ &\multicolumn{2}{|c|}{by KDF} & \multicolumn{2}{c||}{by KFF}& \multicolumn{3}{c|} {by KDF} \\ \cline{4-10}
& &  &$ H$    & $\Delta_{score}$ & $H $   & $\Delta_{score}$ & $ n $    & $H $   & $\Delta_{score}$ \\ \hline
{\tt 1cs0.a} &1073 &0    & 2.67 & N/A              & 2.31 & N/A              &  8 232 & 2.67 & 2.42 \\ \hline
{\tt 1g8k.a} &822  &545  & 2.07 & 4.18             & 1.49 & 4.71             & 11 997 & 2.07 & 1.69 \\ \hline
{\tt 1gqi.a} &708  &1002 & 3.03 & 5.16            & 2.82 & 5.03             & 13 707 & 3.03 & 2.16 \\ \hline
{\tt 1kqf.a} &981  &93   & 2.19 & 5.17             & 1.85 & 4.95             & 9 612  & 2.19 & 1.82 \\ \hline
{\tt 1lsh.a} &954  &148  & 1.97 & 4.57             & 1.66 & 4.02             & 10 017 & 1.97 & 2.01 \\ \hline
\end{tabular}
\end{center}
\end{small}
\end{table}

As any other discrimination problems, the success of classification
strongly depends on the training data.  If the scoring function is
challenged with a drastically different protein than proteins in the
training set, it is possible that the classification will fail.  To
further test how well the nonlinear scoring function performs when
discriminating proteins that are dissimilar to those contained in the
training set, we take five proteins that are longer than any training
proteins (lengths between 46 and 688).  These are obtained from the
list of 1,261 polypeptide chains contained in the updated Oct 15, 2002
release of {\sc Whatif} database.  The first test is to discriminate
the 5 proteins from 1,728 exhaustively generated design decoys using
gapless threading.  The second test is to discriminate these 5
proteins from exhaustively enumerate sequence decoys generated by
threading 14 large protein sequences of unknown structures obtained
from SwissProt database, whose sizes are between 1,124 and 2,459.
This is necessary since structures of the longest chains otherwise
have few or no threading decoys.  Tabel~\ref{tab:largeProteins} lists
results of these test, including the predicted score value and the
smallest gap between the native protein and decoys. For the first
test, the nonlinear design scoring functions can discriminate these 5
native proteins from all decoys in the first test.  For the second
test, the design scoring function can also discriminate all 5 proteins
from a total of 53,565 SwissProt sequence decoys, and the smallest
score gaps between native and decoys are large.

\begin{figure}[t!]
\centerline{\epsfig{figure=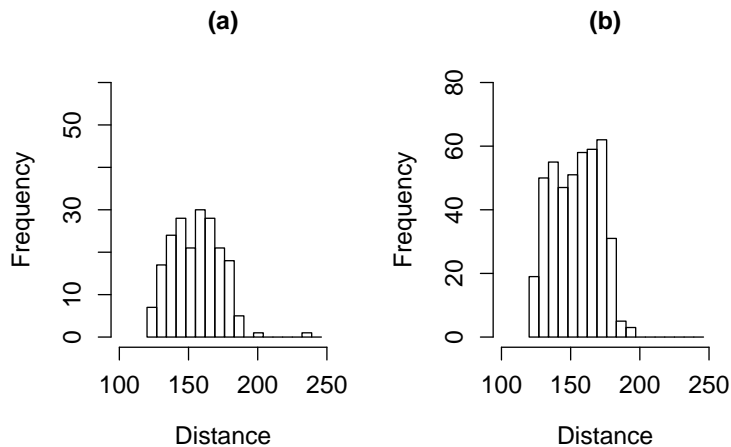,width=4in}}
\caption{\sf
The distribution of maximum distances
of proteins to the set of training proteins.  (a). The maximum distance
for each training protein to all other 439 proteins.  (b). The maximum
distance for each protein in the 201 test set to all 440 training
proteins.  These two distributions are similar.
}
      \label{Fig:maxDist}
\end{figure}

We found that it is infrequent for an unknown test protein to have low
similarity to all reference proteins.  For each protein in the 440
training set, we calculate its Euclidean distance to the other 439
proteins.  The distribution of the 440 maximum distances for each
training protein to all other 439 proteins are shown in
Figure~\ref{Fig:maxDist}a.  We also calculate for each protein in the
201 test set its maximum distance to all training proteins
(Figure~\ref{Fig:maxDist}b).  It is clear that for most of the 201
test proteins, the values of maximum distances to training proteins
are similar to the values for training set proteins.  The only
exceptions are two proteins, ribonuclease inhibitor ({\tt 1a4y.a}) and
formaldehyde ferredoxin oxidoreductase ({\tt 1b25.a}).  Although they
are correctly classified, the former has significant amount of
unaccounted interchain contact with another protein angiogenin, and
the latter has iron/sulfur clusters.  It seems that the set of
training proteins provide an adequate basis set for characterizing the
global fitness landscape of sequence design for other proteins.

\paragraph{Nature of Misclassification.}

\begin{table}[!t]
\begin{small}
\begin{center}
\caption{\sf
The nearest neighbors of the 13
proteins misclassified  by  design function.  The number of
native protein support vectors among the top 3, 5, and 11 nearest
neighbors (NNs) are listed.  Except  protein {\tt 1bx7}, the
majority of nearest neighbors of all misclassified proteins are
decoys.
}
\label{tab:NN}
\vspace*{.3in} 
\begin{tabular}{|c|c|c|c|}
\multicolumn{4}{c}{}\\ \hline
 pdb & 3-NN & 5-NN & 11-NN\\ \hline
 1bd8  &  0 & 0 & 0\\ \hline
 1bx7  & 2 & 3 & 5 \\ \hline
 1bxy.a  & 0  & 1 & 1\\ \hline
 1cku.a  & 1 & 2 & 2\\ \hline
 1dpt.a  & 2 & 2 & 2 \\ \hline
 1flt.v  & 1  & 3 & 3 \\ \hline 
 1hta  &  0 &0  &0 \\ \hline
 1mro.c  &0  &0  &0 \\ \hline
 1ops  & 1 & 1 & 1\\ \hline
 1psr.a  & 1 &1  &1 \\ \hline
 1rb9 & 1 & 1 &1 \\ \hline
 1ubp.b & 1 &1  &1 \\ \hline
 3ezm.a & 0  & 0 & 0\\ \hline
 \end{tabular}
\end{center}
\end{small}
\end{table}

We further distinguish misclassifications due to native protein being
too close to a decoy and misclassifications due to decoys being too
close to a native protein.  Among the set of 201 test proteins, the
native sequences of 13 proteins are not recognized correctly from
design decoys.  These 13 proteins are truly misclassifications because
they do not have extensive unaccounted interchain interactions or
cofactor interactions.  We calculate the Euclidean distance of each of
the 13 proteins from the 220 native protein and 1,685 decoys that
participate in the kernel design scoring function.  The results are
shown in Table~\ref{tab:NN}, where the number of native proteins among
the top 3, 5, and 11 nearest neighboring vectors to the failed protein
are listed.  Except protein {\tt 1bx7}, all misclassifications are due
to native vectors being too close to decoys.

\section{Discussion}

\paragraph{Formulation of Non-linear Scoring Function.}
A basic requirement for computational studies of protein design is an
effective scoring function, which allows searching and identifying
sequences adopting the desired structural templates.  Our study
follows earlier works such as
\citep{Vendruscolo00_Proteins.pair,TobiElber00_Proteins,Goldstein92_PNAS},
where empirical scoring functions based on coarse residue level
representation have been developed by optimization.  The goal of this
study is to explore ways to improve the sensitivity and/or specificity
of discrimination.

There are several routes towards improving empirical scoring
functions.  One approach is to introduce higher order interactions,
where three-body or four-body interactions are explicitly incorporated
in the scoring function
\citep{Zheng97,Munson97,Betancourt99_PS,Rossi02_BJ,LiHuLiang03_Proteins}.
A different approach is to introduce nonlinear terms.  Recently, Fain
{\it et al} uses sums of Chebyshev polynomials upto order 6 for
hydrophobic burial and each type of pairwise interactions
\citep{Fain02_PS}.

In this work, we propose a different framework for developing
empirical protein scoring functions, with the goal of simultaneous
characterization of fitness landscapes of many proteins.  We use a set
of Gaussian kernel functions located at both native proteins and
decoys as the basis set. Decoy set in this formulation are equivalent
to the reference state or null model used in statistical
potential. The expansion coefficients $\{ \alpha_N \}, N \in {\cal N}$
and $\{ \alpha_D \}, D \in {\cal D}$ of the Gaussian kernels determine
the specific form of the scoring function.  Since native proteins and
decoys are non-redundant and are represented as unique vectors $\bc
\in \real^d$, the Gram matrix of the kernel function is full-rank.
Therefore, the kernel function effectively maps the protein space into
a high dimensional space in which effective discrimination with a
hyperplane is easier to obtain.  The optimization criterion here is
not $Z$-score, rather we search for the hyperplane in the transformed
high dimensional space with maximal separation distance between the
native protein vectors and the decoy vectors.  This choice of
optimality criterion is firmly rooted in a large body of studies in
statistical learning theory, where expected number of errors in
classification of unseen future test data is minimized
probabilistically by balancing the minimization of the training error
(or {\it empirical risk}) and the control of the capacity of specific
types of functional form of the scoring function
\citep{Vapnik95,Burges98,ScholkopfSmola02}.

This approach is general and flexible, and can accommodate other
protein representations, as long as the final descriptor of protein
and decoy is a $d$-dimensional vector.  In addition, different
forms of nonlinear functions can be designed using different kernel
functions, such as polynomial kernel and sigmoidal kernels.  It is
also possible to adopt different optimality criterion, for example, by
minimizing the margin distance expressed in 1-norm instead of the
standard 2-norm Euclidean distance.

\paragraph{Folding Scoring Fucntion.}
The geometric views of design scoring function and the
optimality criterion also apply to the protein folding problem.  For
folding scoring function, the only difference from  design
scoring function of Equation (\ref{nonlinear}) is that here $\cal D$
is a set of structure decoys rather than a set of sequence decoys.
Specifically, we generate for each native protein $ (\bs_N, \ba_N)
$ a set of structure decoys $\{ (\bs_D, \ba_N) \}$, {\it i.e.}, by
mounting the native sequence on fragment of the structure of a large
protein such that it contains exactly the same number of amino acid
residues as the native protein.  We use the same training set of 440
protiens from {\sc Whatif98} and 14,080,766 structural decoys as in
design study.  The same optimization technique of margin maximization
is used.  The $\sigma^2$ value and the number of proteins and decoys
entering the final folding scoring function are listed in
Table~\ref{tab:learning}.

For comparison, we also report  discrimination results of the optimal
linear scoring function taken as reported in 
\citep{TobiElber00_Proteins_1}, as well as the statistical potential
developed by Miyazawa and Jernigan.  Here we use the contact
definition reported in \citep{TobiElber00_Proteins_1}, that is, two
residues are declared to be in contact if the geometric centers of
their side chains are within a distance of 2.0 -- 6.4 \AA.

\begin{table}[!t]
\begin{center}
\caption{\small
\sf
The number of misclassified
protein structures for the test set of 194 proteins and the set of 201
proteins using nonlinear kernel folding scoring function, two optimal
linear scoring function taken as reported in
\citep{TobiElber00_Proteins_1}, in Table I of
\citep{Bastolla01_Proteins}, and Miyazawa-Jernigan statistical
potential \citep{Miyazawa96_JMB}. The set of 201 proteins include those
with more than 30\% interchain contacts and those with $>10\%$ missing
coordinates.  We also list performance of kernel design scoring
function for structure recognition.
}
\label{tab:result.fold}
\vspace*{.3in}
\begin{tabular}{|c|c|c|}
\hline
    & { Misclassified Natives} & {Misclassified Natives} \\ \hline \hline
{ Kernel Folding Scoring Function}    &  4/194 & 8/201  \\ \hline
Tobi \& Elber              &  7/194 & 13/201 \\ \hline
Bastolla {\it et al}            &  2/194 & 5/201 \\ \hline 
Miyazawa \& Jernigan            &  85/194& 92/201 \\ \hline
\hline
{ Kernel Design Scoring Function}      &  4/194 &9/201  \\ \hline
\end{tabular}\\
\end{center}
\end{table}

To test nonlinear folding scoring functions for the same 194 and 201
test set proteins, we take the structure $\bs$ from the
conformation-sequence pair $(\bs, \ba_N)$ with the lowest score as the
predicted structure of the native sequence.  If it is not the native
structure $\bs_N$, the discrimination failed and the folding scoring
function does not work for this protein.  The results of
discrimination are summarized in Table~\ref{tab:result.fold}.  There
are 4 and 8 misclassified native structures for the 194 set and 201
set, respectively. These correspond to a failure rate of 2.1\% and
4.0\%, respectively.  The performance of the optimal nonlinear kernel
folding scoring function is better than the optimal linear scoring
function of \citep{TobiElber00_Proteins_1}, based on calculation using
values taken from \citep{TobiElber00_Proteins_1} (failure rates 3.6\%
and 6.5\% for the 194 set and 201 set, respectively), and is
comparable to the results using values taken from reference
\citep{Bastolla01_Proteins} (2 and 5 misclassification, failure rates
of 1.0\% and 2.5\% for the 194 set and 201 set, respectively).
Consistent with previous reports \citep{Clementi98_PRL}, statistical
potential has about $43.8\%$ (81 out of 194) and $43.2\%$ (87 out of
201) failure rates for the 194 set and the 201 set, respectively.

An updated study to reference \citep{Vendruscolo00_Proteins.pair} reported
perfect discrimination for 1,000 proteins from folding decoys \citep{Bastolla01_Proteins}.
Our results cannot be directly compared with this study, because many
of the test proteins or their homologs in our study are likely to be
included in the training set of \citep{Bastolla01_Proteins}, as it is
the union of proteins in the {\sc Whatif} database and the {\sc
Pdbselect} database.  In addition, it is not clear whether all decoys
generated by gapless threading were tested in reference
\citep{Bastolla01_Proteins}.  This makes a direct comparison of the two
studies rather difficult.

It is informative to examine the four misclassified proteins by the
kernel folding scoring function ({\tt 1bx7}, {\tt 1hta}, {\tt 1ops},
and {\tt 3ezm.a}).  Hirustasin {\tt 1bx7} contain five disulfide
bonds, which are not modeled explicitly by the protein description.
{\tt 1hta} (histone Hmfa) exists as a tetramer in complex with DNA
under physiological condition. Its native structure may not be the
same as that of a lone chain. The two terminals of this protein are
rather flexible, and their conformations are not easy to determine.
Among the 13 native sequences misclassified by the kernel design
scoring function ({\tt 1bd8},{\tt 1bx7}, {\tt 1bxy.a}, {\tt 1cku.a},
{\tt 1dpt.a}, {\tt 1flt.v}, {\tt 1hta}, {\tt 1mro.c}, {\tt 1ops}, {\tt
1psr.a}, {\tt 1rb9}, {\tt 1ubp.b}, {\tt 3ezm.a}), several have
extensive interchain interactions, although the contact ratio is below
the rather arbitrary threshold of 30\%: Contact Ratio of $24\%$ for
{\tt 1mor.c}, $19\%$ for{\tt 1upb.b}, $24\%$ for {\tt 1flt.v}, $15\%$
for {\tt 1psr.a}, and $13\%$ for {\tt 1qav.a}.  It is likely that the
substantial contacts with other chains would alter the confirmation of
a protein.  {\tt 1cku.a} (electron transfer protein) contains an
iron/sulfur cluster, which covalently bind to four Cys residues and
prevent them from forming 2 disulfide bonds. These covalent bonds are
not moldeled explicitly.  {\tt 1bvf} (oxidoreductase) is complexed
with a heme and an FMN group. The conformations of {\tt 1cku.a} may be
different upon removing of these functionally important hetero
groups. Altogether, there are some rationalization for 8 of the 13
misclassified proteins.

In many cases, the misclassification of some native conformations are
often indicative of the peculiar nature of the protein structures.
This is true for both linear scoring function reported in
\citep{Vendruscolo98_JCP,Vendruscolo00_Proteins.pair} and the nonlinear
kernel function developed in this study.  For example, the
misclassified proteins are often peptide chains stabilized by other
chains, or by interactions with cofactors, or are small fragments
whose interactions are modified by crystal lattice interactions, or
are NMR structures which are less compact and less stable than X-ray
structures.  Although in this study we attempted to alleviate such
complications by eliminating very short peptide fragments and
excluding proteins with over 30\% interchain contacts, it is unlikely
all problematic protein structures can be completely eliminated from
the training set.  As shown by Bastolla {\it et al\/} in
\citep{Bastolla01_Proteins}, the design of optimized scoring function
is likely to be open to the presence of wrong samples when a large
training set is used.

\begin{figure}[!t]
      \centerline{\epsfig{figure=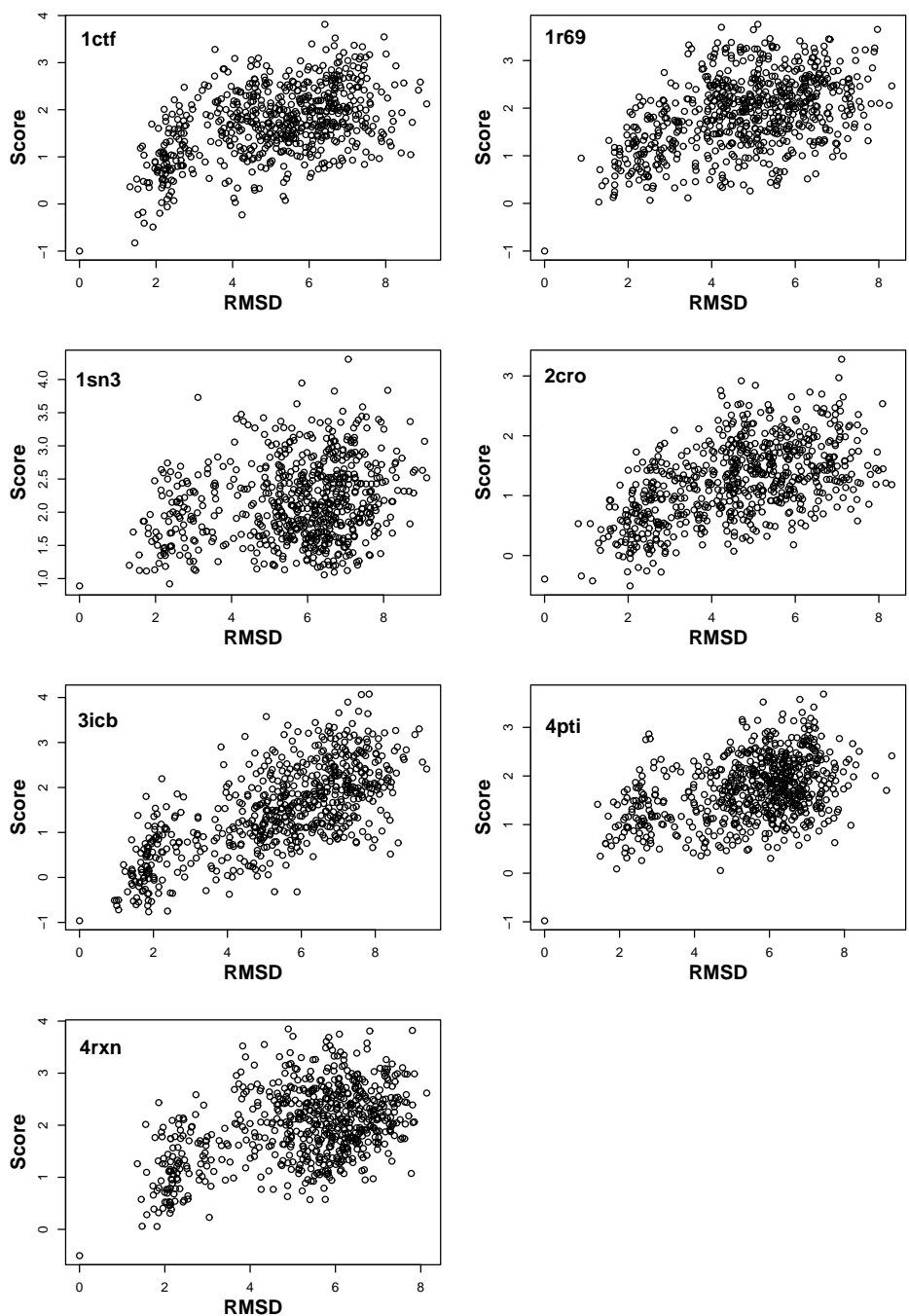,width=5in}}
\caption{\sf
 Correlation of scores of decoys
evaluated  by nonlinear kernel folding scoring function
and their RMSD values to the native proteins in the {\sc
4state\_reduced} set.
}
      \label{Fig:4state}
\end{figure}

For protein folding scoring functions derived from simple decoys
generated by gapless threading, a more challenging test is to
discriminate native proteins from an ensemble of explicitly generated
three dimensional decoy structures with a significant number of
near-native conformations
\citep{ParkLevitt96_JMB,SamudralaMoult98_JMB}.  Here we evaluate the
performance of nonlinear scoring functions using three decoy sets from
the database ``{\sc Decoys 'R' Us}'' \citep{SamudralaLevitt00_PS}: the
{\sc 4state\_reduced} set, the {\sc Lattice\_ssfit} set, and the {\sc
lmsd} set.
We compare our results in performance with results reported in
literature using optimal linear scoring function
\citep{TobiElber00_Proteins} and statistical potential
\citep{Miyazawa96_JMB} (Table~\ref{tab:4state}).  For the {\sc
4state\_reduced} set of decoys, nonlinear folding scoring function has
the best performance in terms of identifying the native structure,
with only one misclassification ({2cro}).  The correlation of root
mean square distance (RMSD) of conformations to the native structure
and score value in the {\sc 4state} set are shown in
Fig~\ref{Fig:4state}.  Although the performance of discriminating
explicit generated challenging decoys is not as good as that of
discriminating decoys generated by threading, it is likely that
nonlinear kernel scoring functions can be further improved if more
realistic structural decoys are included in training.  The
generation of realistic structural decoys is more involved.
Several methods have been developed for generating realistic decoys,
including the original ``build-up'' method \citep{ParkLevitt96_JMB}, 
those with additional energy minimization \citep{Floudas03_Proteins}, and
method based on fragment assembly \citep{Simons97_JMB}.  In additoin,
effective strategy of sequential
importance sampling has also been proposed to generate protein-like
long chain compact self-avoiding walk to overcome the attribution
problem \citep{ZCTL03-jcp}.  This approach has been applied to generate
realistic decoys. Preliminary results of deriving scoring funciton
using such decoys can be found in \citep{Zhang04-EMBC}.

\begin{table}[t!]
\begin{small}
\begin{center}
\caption{
\sf
Results of discrimination of native
structures from decoys using nonlinear kernel scoring functions.  The
decoy sets include  {\sc 4state\_reduced} set, {\sc
Lattice\_ssfit} set, and  {\sc lmsd} set
\citep{SamudralaLevitt00_PS}.
The rank of the native structure and its $z$-score are
listed.  The correlation coefficient $R$ is also listed in parenthesis
for the {\sc 4state\_reduced} set.  KFF stands for kernel folding
scoring function, and KDF stands for kernel design scoring
function. TE-13 scoring function is linear distance based scoring
function optimized by linear programming, taken as reported in
\citep{TobiElber00_Proteins}, BFKV the linear scoring function reported
in \citep{Bastolla01_Proteins}, and MJ is the statistical scoring
function as reported in \citep{Miyazawa96_JMB}.  Results for TE-13
scoring function and Miyazawa-Jernigan scoring function are taken from
Table II of \citep{TobiElber00_Proteins}.
}
\label{tab:4state}
\vspace*{.2in}
\begin{tabular}{ccccccc}
\multicolumn{7}{l}{1. {\sf 4state\_reduced} } \\ \hline
Protein & \# of decoys &  KFF & KDF  &   MJ   &   TE-13  & BFKV \\ \hline
 {\tt 1ctf }  & 631 &  1/3.64(0.49) & 1/3.14(0.55)   &  1/3.73 & 1/4.20&   2/3.00   \\
 {\tt 1r69 }  & 676 &  1/3.77(0.45) & 1/3.79(0.55)   &  1/4.11 & 1/4.06  & 1/4.30  \\
 {\tt 1sn3 }  & 661 &  1/2.15(0.24) & 27/1.79(0.41)   &  2/3.17 & 6/2.70 &  1/2.89   \\
 {\tt 2cro }  & 675 &  3/2.57(0.54 &  1/2.66(0.61)    &  1/4.29 & 1/3.48 &  2/2.91  \\
 {\tt 3icb }  & 654 &  1/2.56(0.70) & 1/2.68(0.74)  &    ---     &    ---      &  1/2.96  \\
 {\tt 4pti }  & 688 &  1/4.17(0.41) &  1/2.79(0.54)   &  3/3.16 & 7/2.43 &  1/3.49  \\
 {\tt 4rxn }  & 678 &  1/3.45(0.47) &  7/1.99(0.53)   &  1/3.09 & 16/1.97&  1/3.32  \\
  \hline
  \multicolumn{7}{l}{2. {\sf lattice\_ssfit}} \\ \hline
  Protein & \# of decoys & KFF &  KDF &    MJ  & TE-13& Bastolla \\ \hline
 {\tt 1beo   } &   2001      & 15/2.45  & 1/3.94 &   ---   & ---      & 1/3.70  \\
 {\tt 1ctf   } &   2001      & 1/3.76  & 1/5.35  &  1/5.35 &  1/6.17  & 1/4.66 \\
 {\tt 1dkt  }  &   2001      & 17/2.42  & 8/2.64  &  32/2.41&  2/3.92  & 4/3.38 \\
 {\tt 1fca }   &   2001      & 56/2.00  & 98/1.76 &  5/3.40 &  36/2.25 & 14/2.56 \\
 {\tt 1nkl  }  &   2001      & 1/3.60  & 1/3.51  &  1/5.09 &  1/4.51   & 1/4.53 \\
 {\tt 1pgb  }  &   2001      & 1/3.95  & 1/4.91  &  3/3.78 &  1/4.13   & 1/3.41 \\
 {\tt 1trl  }  &   2001      & 56/1.97 & 18/2.67&  4/2.91 &  1/3.63   &  90/1.75 \\
 {\tt 4icb  }  &   2001      & 1/3.92  & 1/5.31  &    ---  &   ---   &  1/4.39 \\ \hline
  
  \multicolumn{7}{l}{3. {\sf 1msd}} \\ \hline
  Protein & \# of decoys & KFF &  KDF &    MJ  & TE-13 & Bastolla\\ \hline
 {\tt 1b0n-B } &  498  &  406/-0.94   &  19/2.05  & ---  & ---  & 257/-0.03 \\
 {\tt 1bba  }  &  501  &  500/-3.58   &  487/-1.83 & --- & --- & 500/-3.31 \\
 {\tt 1ctf  }  &  498  &  1/3.62    &  1/3.31    & 1/3.86   & 1/4.13  & 1/2.92 \\
 {\tt 1dtk  }  &  216  &  59/0.64  &  185/-1.11 & 13/1.71  & 5/1.88  & 54/0.74 \\
 {\tt 1fc2  }  &  501  &  501/-3.08 &  486/-1.87 & 501/-6.24& 14/2.04  & 501/-3.84 \\
 {\tt 1igd  }  &  501  &  1/5.18    &  1/3.93    & 1/3.25   & 2/3.11  & 6/2.68 \\
 {\tt 1shf-A}  &  438  &  5/2.14    &  12/1.82    & 11/2.01  & 1/4.13  & 1/3.28 \\
 {\tt 2cro  }  &  501  &  2/2.65    &  1/3.24    & 1/5.07   & 1/3.96  & 1/4.59 \\
 {\tt 2ovo  }  &  348  &  1/3.11    &  38/1.21   & 2/3.25  &  1/3.62 & 40/1.15\\
 {\tt 4pti }   &  344  &  1/3.14    &  108/0.62   & ---  & ---  & 10/1.86 \\
\hline
\end{tabular}
\end{center}
\end{small}
\end{table}

\paragraph{Nonlinear Scoring  Function for Folding and Design.}
      \begin{figure}[t!]
      \centerline{\epsfig{figure=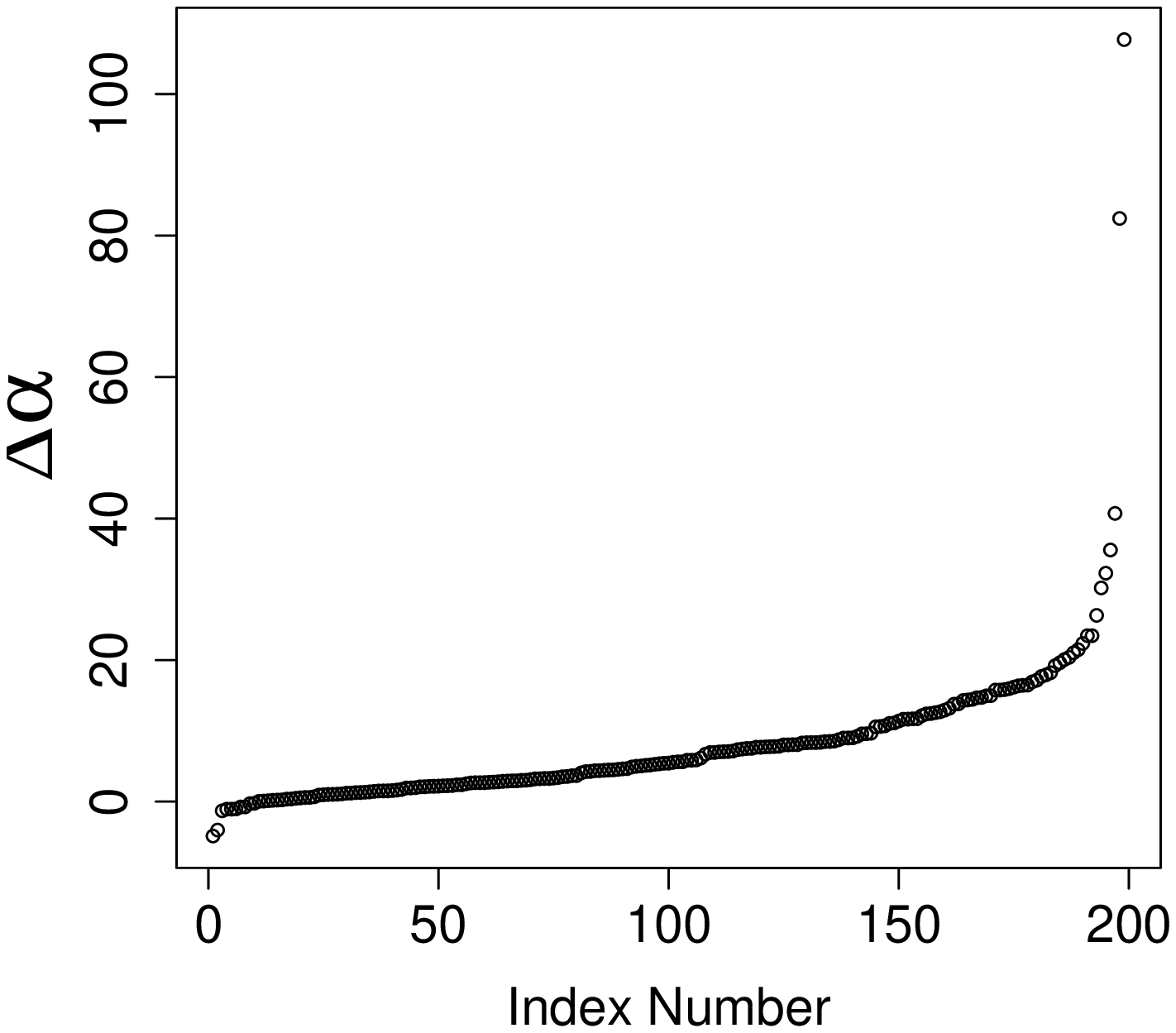,width=3in}}
\caption{\sf
The difference in contribution to
the scoring function for the 199 native protein structures that
participate in both folding and design scoring functions.  They are
sorted by $\Delta \alpha = \alpha_{\mbox{design}} -
\alpha_{\mbox{folding}}$.  The majority of them have $\Delta \alpha$
close to 0.
}
      \label{Fig:alphaDiff}
\end{figure}

\begin{figure}[t!]
      \centerline{\epsfig{figure=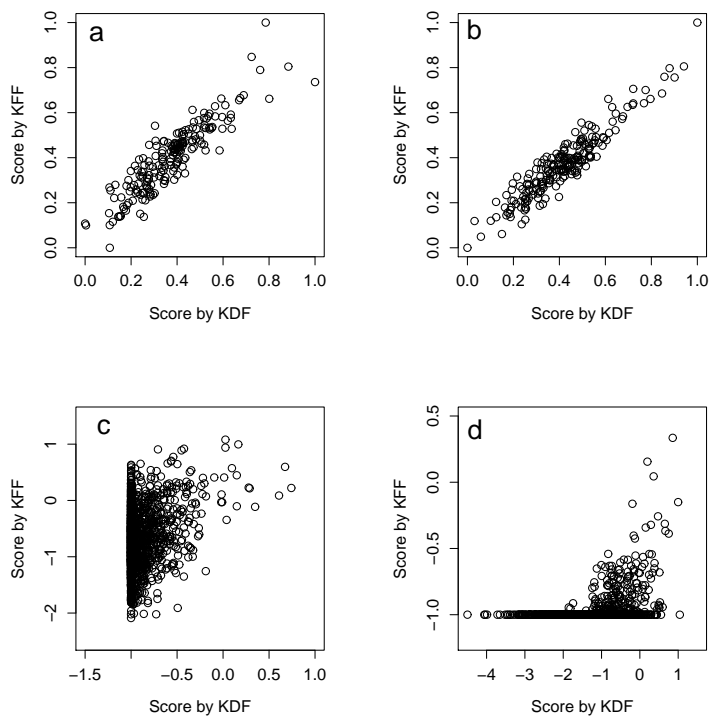,width=4in}}
\caption{\sf
 Comparison of kernel design
scoring function (KDF) and nonlinear kernel folding scoring function
(KFF).  (a). The score values by KFF and by KDF for the 194 proteins
are strongly correlated.  The correlation coefficient is $R = 0.90$.
(b). The score values of the nonlinear design and folding scoring
functions for the 210 unit vectors are strongly correlated ($R =
0.94$).  (c). The score values by both design scoring functions and by
folding scoring functions for decoys that enter the nonlinear design
functions are poorly correlated. (d). The score values for decoys that
enter the nonlinear folding scoring functions are also poorly correlated.
}
      \label{Fig:testSetenergies}
\end{figure}

Sequence decoys and structure decoys in general lead to different
scoring functions.  For example, the contact count vectors $\bc$ can
be very different for a sequence decoy of a protein and a structure
decoy of the same protein.  The discrimination surface defined by the
design scoring function and the folding scoring function therefore may
be different.  There are 220 out of 440 native proteins participating
in design scoring function, and 214 out of 440 native proteins
participating in folding scoring function.  There are 199 proteins
that appear both in folding and design scoring functions.  The
majority of the native proteins have similar $\alpha$ values for both
folding and design scoring functions.  Fig~\ref{Fig:alphaDiff} shows
the difference $\Delta \alpha_i$ of the coefficient $\alpha_i$ for
protein $i$ appearing in both folding scoring function and design
scoring function. In most cases, $\Delta \alpha_i $ values are small.
That is, most native proteins contribute similarly in design scoring
function and in folding scoring function.  This is expected, because
the main differences between the two scoring functions are due to
differences in decoys.  Out of the top 20 proteins with the largest
$|\alpha_i|$ values, 11 are common for both folding and design scoring
functions.  It is possible that the score values by kernel folding
scoring function and by kernel design scoring function may be similar
for many structure-sequence pairs $(\bs, \ba)$.
Figure~\ref{Fig:testSetenergies}a shows that the 194 proteins in the
test set have similar score values by the kernel folding and kernel
design scoring functions.

We also compare the value of the scoring functions for each of the 210
unit vector $\bc_1 = \{1, 0, \ldots, 0\}^T, \cdots$, and $ \bc_{210} =
\{0, \ldots, 1\}^T$. We normalize these values so $\max H(\bc_i) = 1$
for both scoring functions (Fig~\ref{Fig:testSetenergies}b).  There is
strong correlation ($R = 0.94$) for folding and design scoring
functions.

However, other methods reveal that kernel folding and design scoring
functions are different.  One method is to compare the scores of a
subset of decoy structures that are challenging. That is, we compare
evaluated scores of decoys with $\alpha_i \ne 0$.
Fig~\ref{Fig:testSetenergies}c shows that for decoys appearing in the
design scoring functions, there is little correlation in scores
calculated by design scoring function and by folding scoring function.
Similarly, there is no strong correlation between scores calculated by
folding scoring function and by design scoring function for the set of
structure decoys entering the design scoring function
(Fig~\ref{Fig:testSetenergies}d).  It seems that although the values of
$\alpha_N$s are similar for the majority of the native proteins,
design scoring function and folding scoring function can give very
different score values for some conformations.  This suggests that the
overall fitness for design and folding potential may be different.
However, since all empirical scoring functions derived from
optimization and protein structures depend on the choice of traning
set proteins and decoys, we cannot rule out the alternative
explanation that the observed difference between design and folding
scoring functions may be due to the difference of the decoy sets.

\begin{table}[htb]
\begin{small}
\begin{center}
\caption{
\sf
The top twenty proteins with the
largest $\alpha$ value among 199 proteins entering both kernel folding
scoring function and kernel design scoring function.  The $\alpha$
value, the protein class as defined by SCOP, and the number of
residues are also listed.
}
\label{tab:highest_alpha}
\vspace*{.3in}
 \begin{tabular}{l|lccr|lccr}\hline
&\multicolumn{4}{c|}{Kernel Design Scoring Function}&\multicolumn{4}{c}{Kernel Folding Scoring Function} \\ \cline{2-9}
 Index  & pdb & $\alpha$  & Class & Number &   pdb & $\alpha$ & Class & Number\\
   &  &value  & & of resides &    &value  &  &  of resides  \\ \hline
1 &{\tt 2por } & 130.88 & Membrane/cell & 301 &{\tt 2spc.a } & 25.95 &  $ \alpha $ & 107 \\
2 &{\tt 1prn } & 96.73 & Membrane/cell & 289 &{\tt 2por } & 23.19 & Membrane/cell & 301 \\
3 &{\tt 2spc.a } & 52.27 &  $ \alpha $ & 107 &{\tt 1prn } & 14.31 & Membrane/cell & 289 \\
4 &{\tt 1nsy.a } & 51.41 & $ \alpha / \beta $ & 271 &{\tt 1rop.a } & 13.28 &  $ \alpha $ & 56 \\
5 &{\tt 3pch.m } & 45.22 &  $ \beta $ & 236 &{\tt 2wrp.r } & 11.41 &  $ \alpha $ & 104 \\
6 &{\tt 1bkj.a } & 40.37 & $ \alpha + \beta $ & 239 &{\tt 1nsy.a } & 10.68 & $ \alpha / \beta $ & 271 \\
7 &{\tt 1xjo } & 36.02 & $ \alpha / \beta $ & 276 &{\tt 1apy.a } & 10.12 & $ \alpha + \beta $ & 161 \\
8 &{\tt 1bdb } & 34.26 & $ \alpha / \beta $  & 276 &{\tt 1tgs.i } & 9.83 & Small & 56 \\
9 &{\tt 1ppr.m } & 31.70 &  $ \alpha $ & 312 &{\tt 3pch.m } & 9.66 &  $ \beta $ & 236 \\
10 &{\tt 1fiv.a}  & 27.48 &  $ \beta $ & 113 &{\tt 1dan.l}  & 8.80 & Small & 132 \\
11 &{\tt 1hcz } & 27.23 &  $ \beta $ & 250 &{\tt 7ahl.a } & 8.78 & Membrane/cell & 293 \\
12 &{\tt 1tta.a}  & 27.16 &  $ \beta $ & 127 &{\tt 2ilk } & 8.72 &  $ \alpha $ & 155 \\
13 &{\tt 7ahl.a } & 26.69 & Membrane/cell & 293 &{\tt 1ppr.m}  & 8.25 &  $ \alpha $  & 312 \\
14 &{\tt 2rhe } & 26.24 &  $ \beta $ & 114 &{\tt 1bkj.a } & 8.09 & $ \alpha + \beta $ & 239 \\
15 &{\tt 3pch.a}  & 26.23 &  $ \beta $ & 200 &{\tt 1cot } & 8.04 &  $ \alpha $ & 121 \\
16 &{\tt 1snc } & 26.10 &  $ \beta $ & 135 &{\tt 1wht.b } & 7.54 & $ \alpha / \beta $ & 153 \\
17 &{\tt 1wht.b}  & 24.69 & $ \alpha / \beta $ & 153 &{\tt 1vps.a } & 7.25 &  $ \beta $ & 285 \\
18 &{\tt 1cot}  & 23.80 &  $ \alpha $ & 121 &{\tt 1vls}  & 7.05 &  $ \alpha $ & 146 \\
19 &{\tt 1bv1 } & 23.58 &  $ \beta $ & 159 &{\tt 1snc}  & 6.48 &  $ \beta $  & 135 \\
20 &{\tt 2kau.b } & 22.45 &  $ \beta $ & 101 &{\tt 1cmb.a } & 6.48 &  $ \alpha $  & 104 \\
\hline
\end{tabular}
\end{center}
\end{small}
\end{table}

\paragraph{Remarks.}
Our goal in this study is to explore an alternative formulation of
scoring  function and assess the effectiveness of this new approach
with experimental data. The nonlinear scoring functions obtained in
this study should be further improved.  For example, unlike the study
of optimal linear scoring function \citep{TobiElber00_Proteins_1}, where
explicitly generated three-dimensional decoys structures are used in
training, we used only structure decoys generated by threading.  The
test results using the {\sc 4state\_reduced} set and the {\sc
lattice\_ssfit} are comparable or better with other residue-based
scoring function (see Fig~\ref{Fig:4state} and Table~\ref{tab:4state}).  It
is likely that further incorporation of explicit three-dimensional
decoy structures in the training set would improve the protein
scoring function.

The evaluation of the nonlinear scoring function requires more
computation than linear function, but the time require is modest: on
an AMD AThlon MP1800+ machine of 1.54 GHz clock speed with 2 GB
memory, we can evaluate the scoring function for 8,130 decoys per
minute.

Overfitting can be a problem in discrimination.  Overfitting occurs
when the scoring function predicts accurately the outcomes of training
set data, but performs poorly when challenged with unrelated and
unseen test data.  Although our scoring function involves a large
number of basis set proteins and decoys, it does not suffer from
overfitting, because it has good performance in blind test of
discriminating native proteins from both structural and sequence
decoys.

In pursuit of improved sensitivity and specificity in discrimination,
the number of reference decoy and native structures currently entering
the scoring function is large ({\it e.g.}, 1,685 decoys and 220 native
proteins for design scoring function).  However, we expect the scoring
function to be significantly simplified and the number of basis
proteins and decoys reduced considerably.  The use of 1-norm instead
of 2-norm in the objective function of Equation
(\ref{Eqn:PrimalLinear}) will automatically reduce the number of
vectors \citep{ScholkopfSmola02}. In addition, new techniques such as
finite Newton method for reduced support vector machine has recently
shown great promise in further reducing the number of support vectors,
where a reduction ratio of 1\% has been reported \citep{LM01,GM02}.

\paragraph{Conclusion.}
We found in this study that no linear scoring function exists that can
discriminate a training set of 440 native sequence from 14 million
sequence decoys generated by gapless threading.  The success of
nonlinear scoring function in perfect discrimination of this training
set proteins and its good performance in an unrelated test set of 194
proteins is encouraging.  It indicates that it is now possible to
characterize simultaneously the fitness landscape of many proteins,
and nonlinear kernel scoring function is a general strategy for
developing effective scoring function for protein sequence design.

Our study of scoring function for sequence design is a much smaller
task than developing a full-fledged fitness function, 
because we study a restricted version of the protein design problem.
We need to recognize only one sequence that folds into a known
structure from other sequences already known to be part of a different
protein structure, whose identity is hidden during training.  However,
this simplified task is challenging, because the native sequences and
decoy sequences in this case are all taken from real proteins.
Success in this task is a prerequisite for further development of a
full-fledged universal scoring function.  A full solution to the
sequence design problem will need to incorporate additional sequences
of structural homologs as native sequences, as well as additional
decoys sequences that fold into different fold, and decoy sequences
that are not proteins ({\it e.g.}, all hydrophobes).  It is our hope
that the functional form and the optimization technique introduced
here will also be useful for such purposes.

In summary, we show in this study an alternative formulation of
scoring function using a mixture of Gaussian kernels.  We demonstrate
that this formulation can lead to effective design scoring function
that characterize fitness landscape of many proteins simultaneously,
and perform well in blind independent tests.  Our results suggest that
this functional form different from the simple weighted sum of contact
pairs can be useful for studying protein design and protein folding.
This approach can be generalized for any other protein representation,
{\it e.g.}, with descriptors for explicit hydrogen bond and higher
order interactions.

\section{Acknowledgment}
We thank Dr.\ Hao Li for very helpful comments on protein design, Dr.\
Yang Dai for discussion on linear programming, Dr.\ Bosco Ho for
suggestion in presentation, and Dr.\ M\'esz\'aros for help in using
the BPMD package.  Dr.\ Michele Vendruscolo kindly provided the list
of the set of 456 proteins.  This work is supported by funding from
National Science Foundation CAREER DBI0133856, DBI0078270, Office of
Naval Research N000140310329, and National Health Institute NIGMS
GM68958-01.

\section{Appendix}

\begin{lemma} 
For a scoring  function in the form of weighted
linear sum of interactions, a decoy always has score values
higher than the native structure by at least an amount of $b>0$, {\it i.e.},
\begin{equation}
\bw \cdot (\bc_D - \bc_N) > b \quad \mbox{ for all }
\{(c_D - c_N)| D \in {\cal D} \mbox{ and } N \in {\cal N}\}
\label{proof1}
\end{equation}
if and only if the origin $\bf 0$ is not contained within
the convex hull of the set of points $\{ (\bc_D - \bc_N)
| D \in {\cal D} \mbox{ and } N \in {\cal N} \}$.
\end{lemma}

\begin{proof}
Suppose that the origin $\bf 0$ is contained within the convex hull
${\cal A} = \mbox{conv} (\{ \bc_D - \bc_N \})$ of $\{ \bc_D - \bc_N
\}$ and Equation (\ref{proof1}) holds.  By the definition of
convexity, any point inside or on the convex hull $\cal A$ canbe
expressed as convex combination of points on the convex
hull. Specifically, we have:
\[
{\bf 0} = \sum_{(c_D- c_N) \in \cal A} \lambda_{c_D - c_N } \cdot(\bc_D - \bc_N), \quad
\mbox{ and } \quad \sum \lambda_{c_D - c_N } = 1, \lambda_{c_D - c_N} >0.
\]
That is, we have the following contradiction:
\[
0 = \bw \cdot {\bf 0} = \bw \cdot \sum_{c_D- c_N}
 \lambda_{c_D - c_N }
 \cdot(\bc_D - \bc_N) 
  = \sum_{c_D - c_N}
 \lambda_{(c_D, c_N )} 
\cdot \bw \cdot (\bc_D - \bc_w)
> \sum_{c_D- c_N}
 \lambda_{c_D- c_N} 
\cdot b = b.
\]

Because the convex hull can be defined as the intersection of half
hyperplanes derived from the inequalities, if a half hyperplane has a
distance $b>0$ to the origin, all points contained within the convex
hull will be on the other side of the hyperplane \citep{Edels87}.
Therefore, $\bw \cdot (\bc_D - \bc_N) > b$ will hold for all
$\{ (\bc_D - \bc_N) \}$.
\end{proof}

\bibliography{design,nigms,svm,array,prop,pack,prf,lattice,bioshape,liang,potential,career,pd02,pair}
\bibliographystyle{bioinformatics}

\end{document}